\documentclass[reprint,superscriptaddress,amsmath,amssymb,aps,practice]{revtex4-1}
\usepackage{graphicx}
\usepackage{amssymb}
\usepackage{amsmath}
\usepackage{color}
\usepackage{slashed}
\usepackage{ulem}

\usepackage{amsmath,amsfonts,amssymb}

\newcommand{\be}{\begin{equation}}
\newcommand{\ee}{\end{equation}}
\newcommand{\bea}{\begin{eqnarray}}
\newcommand{\eea}{\end{eqnarray}}

\makeatletter
\let\Hy@backout\@gobble
\makeatother

\begin{document}
\title{Entanglement dynamics following a sudden quench: an exact solution}

\author{Supriyo Ghosh}
\email{supriyo.ghosh@saha.ac.in}
\author{Kumar S. Gupta}
\email{kumars.gupta@saha.ac.in}
\affiliation{Theory Division, Saha Institute of Nuclear Physics, HBNI, 1/AF Bidhannagar, Kolkata 700064, India}
\author{Shashi C. L. Srivastava}
\email{shashi@vecc.gov.in}
\affiliation{Variable Energy Cyclotron Centre, HBNI, 1/AF Bidhannagar, Kolkata 700064, India}

\date{\today}

\begin{abstract}

We present an exact and fully analytical treatment of the entanglement dynamics for an isolated system of $N$ coupled oscillators following a sudden quench of the system parameters. The system is analyzed using the solutions of the time dependent Schrodinger's equation, which are obtained by solving the corresponding nonlinear Ermakov equations. The entanglement entropies exhibit a multi-oscillatory behaviour, where the number of dynamically generated time scales increases with $N$. The harmonic chains exhibit entanglement revival and for larger values of $N (> 10)$, we find  near-critical logarithmic scaling for the entanglement entropy, which is modulated by a time dependent factor. The $N=2$ case is equivalent to the two site Bose-Hubbard model in the tunneling regime, which is amenable to empirical realization in cold atom systems.
\end{abstract}

\maketitle

\section{Introduction}

Entanglement is one of the most fundamental yet intriguing features of quantum systems and plays a crucial role in the description of a large
class of quantum phenomena. Various entanglement entropies such as von Neumann \cite{vn,con_wot} and the R\'enyi \cite{rm} have been proposed as measures of entanglement. In a pioneering work, the exact analytical expression for the von Neumann entropy for a time independent Gaussian density matrix was obtained in \cite{sorkin}, which was subsequently generalized for massless free scalar field in \cite{sred}. Since then, various entanglement entropies have found use in probing quantum criticality \cite{eisert,kitaevprl}, distinguishing various phases of topological states of matter \cite{levinprl,balentsnat}, non-equilibrium dynamics \cite{pollmbl,zollerprl} and related phenomena.

The time evolution of entanglement entropy is of great interest in the non-equilibrium dynamics of closed and isolated quantum systems \cite{krishnendu}. Following some initial works \cite{ini1,ini2}, a detailed analysis of the time development of the entanglement entropy in a quantum Ising chain under a quench in the transverse magnetic field was presented in \cite{cala-int}. Subsequently the phenomenon of entanglement dynamics has been investigated in the context of many-body localizations \cite{mb1,mb2,mb3,mb4}, spin chains \cite{sp1} and rings \cite{marino}, diffusive \cite{nonint}, integrable \cite{cala-int,prx}, non-integrable \cite{mkormos} and various other discrete systems \cite{elisa,fabio}.  

One of the main objectives of the present work is to explore the generation and time-dependence of entanglement entropy within an exact and fully analytical framework. For this purpose, we consider a chain of $N$ interacting harmonic oscillators described by the Hamiltonian 
\begin{equation}
\label{h0}
 H=\frac{1}{2}\left[\sum_{j=1}^N(p_j^2+\omega^2 x_j^2)+k\sum_{j=1}^{N-1}(x_j-x_{j+1})^2\right],
\end{equation}
where $j = 1$ to $N$ denotes the number of oscillators and we have set the particles masses to unity. Such a system can be experimentally realized using optical tweezers \cite{science} and the individual coupling parameters can be tuned using ultracold atoms \cite{hunger} or Rydberg states \cite{buchmann2017,macri}. Properties of the reduced density matrix for similar systems \cite{ingo1,ingo2} as well as the relationship of the reduced density matrix with correlation functions \cite{ingo3} has been discussed in the literature in the time independent context. Here we present a time dependent analysis of the reduced density matrix and the entanglement dynamics of the system (\ref{h0}) following a sudden quench of the system parameters, using the exact solutions of the time dependent Schr\"odinger's equation (TDSE) \cite{osc1,osc2}. The time dependence of the wave functions is encoded in the solutions of the corresponding non-linear Ermakov equations \cite{osc2,pinney,demler,delCampo}, which ensures that the dynamics of the entangled state is fully consistent with the TDSE through out the quench protocol.  The Ermakov equations reveal the existence of dynamically generated multiple time scales, which lead to a multi-oscillatory behaviour of the entanglement entropies. The use of the Ermakov equations to describe the time dependence of the entanglement entropy is an important feature of the present work, which distinguishes it from other approaches \cite{unanyan,bianchi1,bianchi2} . It may be noted the certain oscillatory behaviour for the entanglement entropy \cite{mkormos,elisa} in discrete systems have been observed using numerical techniques. The exact analytical treatment presented here provides a more detailed and comprehensive picture of such entanglement oscillations.

We start the discussion with the $N=2$ case, which can be mapped to a two site Bose-Hubbard model with time dependent frequency and coupling. This model describes a double well potential loaded with bosonic particles in the  tunneling regime, which has been realized in the laboratory \cite{albiez}. Next we discuss the entanglement dynamics of full $N$-body system, using the exact solutions of the TDSE and derive the exact analytical expressions for the R\'enyi and von Neumann entanglement entropies. We present the results for the case $N=4,6,10,16,20$ and discuss various physical properties of our system such as the entanglement revival \cite{eisler-z}, near-critical scaling \cite{wilczek,korepin,tonni} and the connection with the area law \cite{eisert}. We conclude the paper with a summary and outlook.


\section{Two site Bose-Hubbard model}
Two site Bose-Hubbard model \cite{campbell} in second quantized form with $\hbar=1$ is
described by the Hamiltonian,
\begin{equation}
\label{h1}
 H=\omega_{BH}(a_1^\dagger a_1+a_2^\dagger a_2)-J(a_1^\dagger a_2 + a_2^\dagger a_1).
\end{equation}
In terms of the canonically conjugate coordinates 
\begin{equation}
 X_j=\frac{a_j+a_j^\dagger}{\sqrt{2}}, ~~~~~~P_j=\frac{a_j-a_j^\dagger}{\sqrt{2}i},
\end{equation}
where $j = 1,2$, the Bose-Hubbard Hamiltonian takes the form
\begin{equation}
\label{h2}
 H=\frac{\omega_{BH}}{2}(X_1^2+P_1^2+X_2^2+P_2^2)-J(X_1X_2+P_1P_2).
\end{equation}
We now define a new set of coordinates $x_\pm$ and momenta $p_\pm$ as
\begin{equation}
\begin{aligned}
 x_+ &\equiv \frac{X_+}{\sqrt{\omega_{BH} - J}}, \quad\quad \quad  x_- \equiv \frac{X_-}{\sqrt{\omega_{BH} + J}} \\
p_+ &\equiv \sqrt{\omega_{BH} - J} P_+, ~~~p_- \equiv \sqrt{\omega_{BH} + J} P_-
\end{aligned}
\end{equation}
where $P_{\pm}=\frac{P_1\pm P_2}{\sqrt{2}}$ and $X_{\pm}=\frac{X_1\pm
X_2}{\sqrt{2}}$. The Hamiltonian $H$ in ($\ref{h2}$) can now be
written as
\begin{equation}
\label{h3}
H=\frac{1}{2} \left [p_+^2 + p_-^2 + \omega_+^2 x_+^2 + \omega_-^2  x_-^2 \right ]
\end{equation}
with $\omega_+ = (\omega_{BH}-J)$ and $\omega_- =
(\omega_{BH}+J)$. Note that $(x_+,p_+)$ and $(x_-,p_-)$ are
canonically conjugate and commute between $+/-$ indices. The
Bose-Hubbard Hamiltonian in ($\ref{h2}$) therefore can be expressed as
two commuting harmonic oscillators with frequencies $\omega_+$ and
$\omega_-$.

Furthermore, it may be noted that the Hamiltonian in ($\ref{h3}$) can
be equivalently expressed as that of two coupled oscillators
\begin{equation}
\label{h4}
 H=\frac{1}{2} \left [ p_1^2+p_2^2+\omega^2(x_1^2+x_2^2)+k(x_1-x_2)^2 \right ]
\end{equation}
with the identifications $p_{\pm}=\frac{p_1\pm p_2}{\sqrt{2}}$ and
$x_{\pm}=\frac{x_1\pm x_2}{\sqrt{2}}$ and with $\omega= \omega_+ =
(\omega_{BH}-J)$ and $\omega_- = \sqrt{\omega^2 + 2k} =
(\omega_{BH}+J)$. These relations imply that the coupling parameter
$k$ in ($\ref{h4}$) is given by $k =2\omega_{BH}J$. This necessitates
the sign of $k$ to be always positive as long as we relate the
($\ref{h4}$) to Bose-Hubbard model. We are thus able to express the
Hamiltonian of the two site Bose-Hubbard model given in ($\ref{h1}$)
as the Hamiltonian of two uncoupled oscillators as in ($\ref{h3}$) or
equivalently as the Hamiltonian of two coupled oscillators as in
($\ref{h4}$) with suitable identification of the respective system
parameters.


\section{Time dependent wave-function for N=2}

In this section we obtain the wave-function of the Bose-Hubbard model
where the parameter $\omega_{BH}$ in ($\ref{h1}$) or equivalently the
parameters $\omega$ and $k$ in ($\ref{h4}$) are explicitly time
dependent. Our strategy would be to solve the TDSE for the the
Hamiltonian in ($\ref{h3}$) using the non-linear Ermakov equation \cite{pinney} and from there to obtain the time
dependent solution for the two coupled harmonic oscillators described
by ($\ref{h4}$).

The TDSE for the Hamiltonian in ($\ref{h3}$) is given by,
\begin{equation}
\label{th3}
\begin{aligned}
 i \frac{\partial \psi (x_+,x_-,t)}{\partial t} &= \frac{1}{2} \left [-\frac{\partial^2}{\partial x_+^2} -\frac{\partial^2}{\partial x_-^2} +
\omega_+^2 x_+^2 + \omega_-^2  x_-^2 \right ]\\ &~\quad \quad \quad\times \psi (x_+,x_-,t),
\end{aligned}
\end{equation}
where $\omega_+(t)=\omega(t)$,
$\omega_-(t)=\sqrt{\omega(t)^2+2k(t)}$. 
Thus the ground state $\psi_0$ of the system at time
$t=0$ is given by
\begin{equation}
\label{gs0}
\begin{aligned}
\psi_{0}(x_+,x_-,t=0) &=\frac{(\omega_+(0)\omega_-(0))^{1/4}}{\sqrt{\pi}}\\
                       &~\times {\rm exp}\Big[-\frac{(\omega_+(0)x_+^2+\omega_-(0)x_-^2)}{2}\Big].
\end{aligned}
\end{equation}
The full time dependent wave function is obtained by solving
($\ref{th3}$), where the initial value of the wave-function is given
in ($\ref{gs0}$). Following the techniques developed in
\cite{osc1,osc2}, full time dependent solution of ($\ref{th3}$) can be
written as 
\begin{align}
 \nonumber\psi(x_+,x_-,t)=&\exp\Big[\frac{im\dot b_1}{2b_1}x_+^2-iE_+\tau_+\Big] \psi(\frac{x_+}{b_1},0)
  \\&\times\exp\Big[\frac{im\dot b_2}{2b_2}x_-^2-iE_-\tau_-\Big] \psi(\frac{x_-}{b_2},0),
\end{align}
where $E_{\pm}$ are the energies of the two decoupled systems at time $t=0$ with $\tau_+=\int_0^t\frac{dt^\prime}{b_1^2(t^\prime)}$,
$\tau_-=\int_0^t\frac{dt^\prime}{b_2^2(t^\prime)}$ and $b_1(t)$, $b_2(t)$ are the scaling parameters which satisfy
the nonlinear Ermakov equations \cite{osc2,pinney} 
\begin{equation}\label{eq:ermakov}
\begin{aligned}
\ddot b_1+\omega_+^2(t)b_1=\frac{\omega_+^2(0)}{b_1^3} ~~{\mathrm{and}}~~\ddot b_2 +\omega_-^2(t)b_2 = \frac{\omega_-^2(0)}{b_2^3} .
\end{aligned}
\end{equation} 
In terms of the coordinates $x_1$, $x_2$ appearing in (\ref{h4}), the wavefunction takes the form
\begin{equation}\label{sol2}
\begin{aligned}
\psi(x_1,x_2,t)=&\tilde{A}(t){\exp}\Big [i\Big(a_1x_1^2
 +a_1x_2^2+2a_2x_1x_2\Big)\Big]
\\&~~~\times {\exp}\Big [-i\Big(E_+\tau_++E_-\tau_-\Big)\Big]
 \\&~~~\times {\exp}\Big[-\frac{1}{4b_1^2}\omega_+(0)(x_1+x_2)^2\Big]\\&~~~\times{\exp}\Big[-\frac{1}{4b_2^2}\omega_-(0)(x_1-x_2)^2\Big].
\end{aligned}
\end{equation}
where $\tilde{A}(t)=\frac{(\omega_+(0)\omega_-(0))^{1/4}}{\sqrt{\pi b_1(t)b_2(t)}}$, $a_1(t)=(\frac{\dot b_1}{4b_1}+\frac{\dot b_2}{4b_2})$, $a_2(t)=(\frac{\dot b_1}{4b_1}-\frac{\dot b_2}{4b_2})$. Note that throughout the text, the quantities $\omega_{\pm}(0)$ correspond to their values just before the quench.


Using ($\ref{sol2}$), the density matrix can be written as 
\begin{align}
\label{density}
\rho(x_1,x_2,x_1^\prime,x_2^\prime,t)=\psi(x_1,x_2,t)\psi^*(x_1^\prime,x_2^\prime,t).
\end{align}


\section{Entanglement entropy}
The R\'enyi entropy of order $\alpha$ is defined as 
\begin{equation}
 S_\alpha=\frac{1}{1-\alpha}\log {\text {Tr}}(\rho_{red}^\alpha),
\end{equation}
where $\alpha$ is any positive integer. The von Neumann entropy can be
obtained in the limit $\alpha \rightarrow 1$. The reduced density matrix is defined as
\begin{equation}
\label{red}
 \rho_{red}(x_1,x_1^\prime,t)=\int dx_2 \rho(x_1,x_2,x_1^\prime,x_2,t).
\end{equation}
Using ($\ref{density}$) and ($\ref{red}$) we get
\begin{equation}\label{red2}
\begin{aligned}
 \nonumber \rho_{red}
(x_1,x_1^\prime,t) = &\pi^{-1/2}(\gamma-\beta)^{1/2}
\exp\left[i(x_1^2-x_1^{\prime2})z(t)\right. \\ 
&~\quad \quad \quad \left. -\frac{\gamma}{2}(x_1^2+x_1^{\prime 2})
 +\beta x_1x_1^\prime \right]
\end{aligned}
\end{equation}
where
\begin{equation}
\begin{aligned}
&\gamma =\frac{\left(\frac{\omega_+(0)}{b_1^2(t)}+\frac{\omega_-(0)}{b_2^2(t)}\right)}{2}-\frac{\left(\frac{\omega_+(0)}{b_1^2(t)}-\frac{\omega_-(0)}{b_2^2(t)}\right)^2-
\left(\frac{\dot 
b_1}{b_1}-\frac{\dot b_2}
{b_2}\right)^2}{4\left(\frac{\omega_+(0)}{b_1^2(t)}+\frac{\omega_-(0)}{b_2^2(t)}\right)},\\
&\beta =\frac{\left(\frac{\omega_+(0)}{b_1^2(t)}-\frac{\omega_-(0)}{b_2^2(t)}\right)^2+\left(\frac{\dot
b_1}{b_1}-\frac{\dot b_2}
{b_2}\right)^2}{4\left(\frac{\omega_+(0)}{b_1^2(t)}+\frac{\omega_-(0)}{b_2^2(t)}\right)},\\
&z(t) =\left(\frac{\dot
b_1}{4b_1} +\frac{\dot
b_2}{4b_2}\right)-\frac{\frac{\omega_+(0)}{b_1^2(t)}-\frac{\omega_-(0)}{b_2^2(t)}}{\frac{\omega_+(0)}{b_1^2(t)}+\frac{\omega_-(0)}{b_2^2(t)}}\left(\frac{\dot
b_1}{4b_1}-\frac{\dot b_2}{4b_2}\right).
\end{aligned}
\end{equation}
To calculate
various entropies, we first obtain the eigenvalues of the reduced
density matrix from the equation,
\begin{equation}
\label{redev}
 \int_{-\infty}^\infty dx_1^\prime \rho_{red} (x_1,x_1^\prime,t) f_n(x_1^\prime,t)=p_n(t)f_n(x_1,t),
\end{equation}
where $n$ is an integer labelling the eigenvalues and the
corresponding eigenfunctions.  The solutions of the eigenvalue
equation ($\ref{redev}$) can be written as
\begin{align} 
&f_n(x_1,t)=H_n(\sqrt{\epsilon}x_1){\rm exp}
\left[-\epsilon \frac{x_1^2}{2} +i x_1^2z(t) \right ],\\
&p_n(t)=(1-\xi(t))\xi(t)^n,
\end{align}
where $H_n$ is the $n^{{\rm th}}$ Hermite polynomial,
$\epsilon=(\gamma^2-\beta^2)^{1/2}$ and explicit time-dependence of $\xi(t)$ is given by, 
\begin{align}\label{eq:xi}
 \xi(t)&=\frac{\beta}{\gamma+\epsilon} =\frac{\frac{\beta}{\gamma}}{1+\sqrt{1-\frac{\beta^2}{\gamma^2}}} < 1,
\end{align}
where $\frac{\beta}{\gamma}=\frac{\left(\frac{\omega_+(0)}{b_1^2(t)}-\frac{\omega_-(0)}{b_2^2(t)}\right)^2+\left(\frac{\dot b_1}{b_1}-\frac{\dot b_2}{b_2}\right)^2}{2\left(\frac{\omega_+(0)}{b_1^2(t)}+\frac{\omega_-(0)}{b_2^2(t)}\right)^2-\left(\frac{\omega_+(0)}{b_1^2(t)}-\frac{\omega_-(0)}{b_2^2(t)}\right)^2+\left(\frac{\dot b_1}{b_1}-\frac{\dot b_2}{b_2}\right)^2}$.
 It may be noted that these
expressions reduce to their time independent counterparts
\cite{sorkin,sred} when the system parameters are time independent.

Thus the R\'enyi entropy can be immediately calculated as,
\begin{equation}
\label{ren}
 S_\alpha(t)=\frac{1}{1-\alpha}\log\frac{(1-\xi(t))^\alpha}{1-\xi(t)^\alpha}.
\end{equation}
The von Neumann entropy can also now be written as
\begin{align}
\label{vn}
 S_1(t)=-\log(1-\xi(t))-\frac{\xi(t)}{1-\xi(t)}\log\xi(t).
\end{align}
 
 This gives the most general form of the entanglement entropies for the arbitrary time dependence in the Hamiltonian parameters.

\section{Quench and entanglement dynamics}
Let us consider the two site Bose-Hubbard model. At at time $t=0$, $\omega_{BH}(t)$ is suddenly quenched from a constant value $\omega_{BH}(i)$ to another constant value $\omega_{BH}(f)$, while the hopping strength  $J$ remains constant. This implies that both the $\omega$ and $k$ are quenched as well. The solutions of the Ermakov equations (\ref{eq:ermakov}) with the boundary conditions $b_{1(2)}(t=0) =1$ and $\dot{b}_{1(2)} (t=0) = 0$, for this quench are given by
\begin{align}
  &b_1(t)=\sqrt{n_1\cos(2(\omega_{BH}(f)-J)t)+m_1},\label{eq:b1gen}\\
 &b_2(t)=\sqrt{n_2\cos(2(\omega_{BH}(f)+J)t)+m_2}\label{eq:b2gen},
\end{align}
where $n_1=\frac{(\omega_{BH}(f)-J)^2-(\omega_{BH}(i)-J)^2}{2(\omega_{BH}(f)-J)^{2}}$, $m_1=\frac{(\omega_{BH}(f)-J)^2+(\omega_{BH}(i)-J)^2}{2(\omega_{BH}(f)-J)^{2}}$ $n_2=\frac{(\omega_{BH}(f)+J)^2-(\omega_{BH}(i)+J)^2}{2(\omega_{BH}(f)+J)^2}$ and
$m_2=\frac{(\omega_{BH}(f)+J)^2+(\omega_{BH}(i)+J)^2}{2(\omega_{BH}(f)+J)^2}$.
Using these expressions of $b_1, b_2$ and Eqns. (\ref{eq:xi}), (\ref{ren}) and (\ref{vn}), we can obtain the various R\'enyi entropies including the von Neumann entropy. 

\begin{figure}[ht!]
\centering
\includegraphics{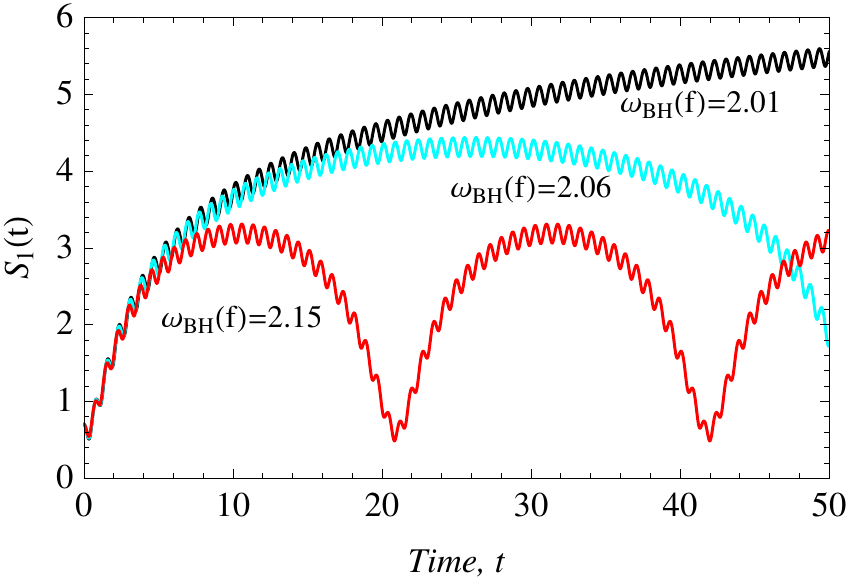}
\caption{Plots of von Neumann entropy for $N=2$. Here $\omega_{BH}(i)=3$ and $J=2$. The $\omega_{BH}(i)$ is quenched to three different values labelled by $\omega_{BH}(f) = 2.15, 2.06$ and $2.01$. When $\omega_{BH}(f)=2.15$, the entropy exhibits a bi-oscillatory behaviour (red line). As the final value of $\omega_{BH}$ approaches $J$, one of the periods of oscillations grows larger, as can be seen for $\omega_{BH}(f)=2.06$ (cyan line). In the limiting case when $\omega_{BH}(f)\rightarrow J$, the overall envelope of the entanglement entropy increasing monotonically with time (black line). }\label{fig:oscl_entropy}
\end{figure}

As an illustration of our results, in Fig. \ref{fig:oscl_entropy} we plot the von Neumann entropy where $\omega_{BH}(i)=3$, $J=2$ and for three values of 
$\omega_{BH}(f)=2.15, 2.06$ and $2.01$. The last value of $\omega_{BH}(f)$ is chosen to investigate the behaviour of the von Neumann entropy in the limit 
$\omega_{BH}(f) \rightarrow J$. The entropy exhibits a bi-oscillatory behaviour with two distinct periods. These periods are same as those appearing in the solutions of the Ermakov equations (\ref{eq:b1gen}) and (\ref{eq:b2gen}). As $\omega_{BH}(f) \rightarrow J$, the time period of the larger of the two oscillations grows further and mimics a polynomial growth at a shorter scale. We emphasize that the multiple time scales appearing here are generated dynamically.

\section{N coupled oscillators}
The Hamiltonian for $N$ coupled oscillators with time dependent parameters is given by
\begin{equation}
\label{H}
\begin{aligned}
H^N(t)&=\frac{1}{2}\left[\sum_{j=1}^N(p_j^2+\omega^2(t)x_j^2)+k(t)\sum_{j=1}^{N-1}(x_j-x_{j+1})^2\right]\\
&=\frac{1}{2}\left[\sum_{j=1}^Np_j^2+X^T.K(t).X\right],
\end{aligned}
\end{equation}
where $X$=$(x_1,x_2...x_N)^T$ and $K$ is a real symmetric $N\times N$ matrix with real eigenvalues. This Hamiltonian can be decoupled using an  orthogonal transformation $U$ to give,
\begin{equation}
 H^\prime=\frac{1}{2}\left[\sum_{j=1}^N(P_j^2+y_j^2K^D_{jj}(t))\right],
\end{equation}
where $Y=UX=(y_1,..,y_N)^T$ and $K^D=UKU^T$ is a diagonal matrix. The time dependent ground state
of the Hamiltonian can be written as
\begin{equation}
\label{Nground}
\begin{aligned}
 &\psi(y_1,..,y_N,t)=\left(\prod_{j=1}^N\frac{1}{b_j^2(t)}\text{det}\frac{\sqrt{K^{D}}}{\pi}\right)^{\frac{1}{4}}\\&~~\times\exp\left[i\sum_{j=1}^N\left(\frac{\dot b_j}{2b_j}
 y_j^2-E_j\tau_j\right)\right]\times\exp\left[\sum_{j=1}^N\frac{y_j^2\sqrt{K^D_{jj}}}{2b_j^2(t)}\right],
 \end{aligned}
\end{equation}
which can be expressed as 
\begin{equation}\label{wf}
\begin{aligned}
 \psi(x_1,....,x_N,t)=&\left(\text{det} \frac{\Omega}{\pi}\right)^{\frac{1}{4}}\exp\left[i\left(X^T\tilde{b}X-\sum_{j=1}^NE_j\tau_j\right)\right]
 \\
 &~~\times\exp\left[-\frac{X^T\Omega X}{2}\right].
 \end{aligned}
\end{equation}
Here $\Omega=U^T\sqrt{K^{\prime D}}U$, $K_{jj}^{\prime D}=\frac{K_{jj}^D(0)}{b_j^4(t)}$, $\tilde{b}=U^T\tilde{b}^DU$ and $\tilde{b}^D$ is a diagonal matrix with elements $\frac{\dot b_j(t)}{2b_j(t)}$. The Ermakov equation satisfied by $b_j(t)$ is  given by
\begin{equation}
\label{4-erm}
 \ddot b_j+\lambda_j^2(t)b_j=\frac{\lambda_j^2(0)}{b_j^3},
\end{equation}
where $\lambda_j$ is the $j$th eigenvalue of the matrix $K(t)$. The time dependent density matrix of the whole system has the form
\begin{equation}\label{den1}
\begin{aligned}
 \rho(X,X^\prime,t)=&\left(\text{det} \frac{\Omega}{\pi}\right)^{\frac{1}{2}}\exp\left[i\left(X^T\tilde{b}X-X^{\prime T}\tilde{b}X^\prime\right)\right]
 \\
 &\exp\left[-\frac{X^T\Omega X}{2}-\frac{X^{\prime T}\Omega X^\prime}{2}\right],
 \end{aligned}
\end{equation}

We partition the whole system in two subsystems $A$ and $B$ with degrees of freedom $n$, $N-n$ and coordinates $\{X^\alpha\}$, $\{X^a\}$ respectively.
Tracing over the subsystem $A$, the reduced density matrix of $B$ is given by
\begin{equation}\label{RD}
 \rho_{red}(X^a,X^{\prime a},t)=\int \prod_{\alpha=1}^n dX^\alpha \rho(X^a,X^\alpha,X^{\prime a},X^\alpha).
\end{equation}
To do this integration we write the matrices $\Omega$ and $b$ respectively as \cite{sorkin,sred},
\begin{equation}
\begin{aligned}
\Omega=
\begin{pmatrix} 
\Omega_{n\times n} & \Omega_{n\times N- n} \\ \Omega_{n\times N-n}^T & \Omega_{N-n\times N-n} 
\end{pmatrix},\\
\tilde{b}=
\begin{pmatrix} 
\tilde{b}_{n \times n} & \tilde{b}_{n \times N-n}\\ \tilde{b}_{n \times N-n}^T & \tilde{b}_{N-n \times N-n} 
\end{pmatrix}.
\end{aligned}
\end{equation}
Using (\ref{den1}) in (\ref{RD}) and after some algebra, we get 
\begin{equation}\label{RD1}
\begin{aligned}
 &\rho_{red}(X^a,X^{\prime a},t)\\~&= \left(\frac{\text{det} \frac{\Omega}{\pi}}{\text{det} \frac{\Omega_{n\times n}}{\pi}}\right)^{\frac{1}{2}}\exp\left[i\left (
X^{aT}ZX^a-X^{\prime aT}ZX^{\prime a}\right )\right]\\&~~\times\exp\left[-\frac{1}{2}\left(X^{a T}\gamma X^a+X^{\prime a T}\gamma X^{\prime a}\right)+X^{aT}\beta X^{\prime a}\right],
 \end{aligned}
\end{equation}
where $X^a$, $X^{\prime a}$ has $N-n$ components and
\begin{equation}
 \begin{aligned}
&Z(t)=\tilde{b}_{N-n \times N-n}-\tilde{b}_{n \times N-n}^T\Omega_{n\times n}^{-1}\Omega_{n\times N- n},\\
&\gamma(t)=\Omega_{N-n\times N-n}-\frac{1}{2}\Omega_{n\times N- n}^T\Omega_{n\times n}^{-1}\Omega_{n\times N- n}\\
&~~~~~~~+2\tilde{b}_{n \times N-n}^T\Omega_{n\times n}^{-1}\tilde{b}_{n \times N-n}, \\
&\beta(t)=\frac{1}{2}\Omega_{n\times N- n}^T\Omega_{n\times n}^{-1}\Omega_{n\times N- n}+2\tilde{b}_{n \times N-n}^T\Omega_{n\times n}^{-1}\tilde{b}_{n \times N-n},
 \end{aligned}
\end{equation}
 which are $(N-n)\times(N-n)$ matrices. 

The reduced density matrix can be written in a product form upto a phase in new coordinates $R^a$ using the orthogonal transformations $V$ and $W$ such that $X=V^T\gamma_D^{-1/2}WR$. Here $V$ is the diagonalizing matrix of $\gamma$ such that $\gamma=V^T\gamma_DV$ and $W$ diagonalizes $\tilde{\beta}$ where $\tilde{\beta}=\gamma_D^{-1/2}V\beta V^T\gamma_D^{-1/2}$. The reduced density matrix in new coordinates takes the form
\begin{equation}
\begin{aligned}\label{RDf}
 &\rho_{red}(R^a,R^{\prime a},t)
=\exp\left[i R^{aT}Z^{\prime\prime} R^a-i R^{\prime aT}Z^{\prime\prime} R^{\prime a}\right]\\&~~~\times\prod_{j=n+1}^N \frac{\left(1-\tilde{\beta_j}\right)^{\frac{1}{2}}}{\pi^{N-n}}\exp\left[-\frac{1}{2}(r_j^2+r_j^{\prime2})+\tilde{\beta_j}r_jr_j^{\prime}\right],
\end{aligned}
\end{equation}
where $R^{a}=(r_{n+1},...,r_N)^T$, $Z^{\prime\prime}=W^T\gamma_D^{-1/2}VZ V^T\gamma_D^{-1/2}W$ and $\tilde{\beta_j}$ are the eigenvalues of $\tilde{\beta}$ matrix. The eigenvalue equation for the reduced density matrix is given by
\begin{equation}
\label{redevN}
 \int_{-\infty}^\infty \mathcal{D}R^{\prime a}\rho_{red}(R^a,R^{\prime a},t)f_l(R^{\prime a},t)=p_l(t)f_l(R^a,t).
\end{equation}
The eigenfunctions are given by
\begin{equation}
 f_l(R^a,t)=H_l(\epsilon^{1/2} R^a)\exp[-R^{aT}\frac{\epsilon}{2}R^a]\exp\left[i R^{aT}Z^{\prime\prime} R^a\right],
\end{equation}
where $H_l(\epsilon^{1/2} R^a)=\prod_{j=n+1}^N H_l(\epsilon^{1/2} r_j)$ and $H_l$ denotes the Hermite polynomial of degree $l$. The eigenvalues are given by $p_l(t)=\prod_{j=n+1}^N (1-\xi_j)\xi_j^l$, where $\xi_j(t)=\frac{\tilde{\beta_j}}{1+\sqrt{1-\tilde{\beta_j}^2}}$. Hence the R\'enyi entropy of order $\alpha$ is given by
\begin{equation}
 S_\alpha(t)=\sum_{j=N-n}^{N}S_{\alpha}[\xi_j(t)],
\end{equation}
where each $S_\alpha[\xi_j(t)]$ has the form as (\ref{ren}) or (\ref{vn}).

\section{Results and physical interpretations}
In order to illustrate the key physical features of the above analysis, we explicitly consider the cases for $N=4,6,10,16$ and 20 for a periodic harmonic chain. We shall focus on how the dynamically generated multiple time scales and the particle number $N$ affect the entanglement dynamics. Note that the time dependence is completely determined by the solutions $b_j$ of the Ermakov eqns. (\ref{4-erm}), which depends on the eigenvalues $\lambda_j$ of the matrix $K$ given by $\lambda_j=\omega^2 + 2k[1-\cos(\frac{2\pi  j}{N})]$, where $j=1,2,...N$. The number of the time scales contributing to the entanglement dynamics for any $N$ depends on the number of distinct eigenvalue of the corresponding $K$. 

  

 \begin{figure}[ht!]
\label{fig2}
\centering
\includegraphics[width=8.6cm]{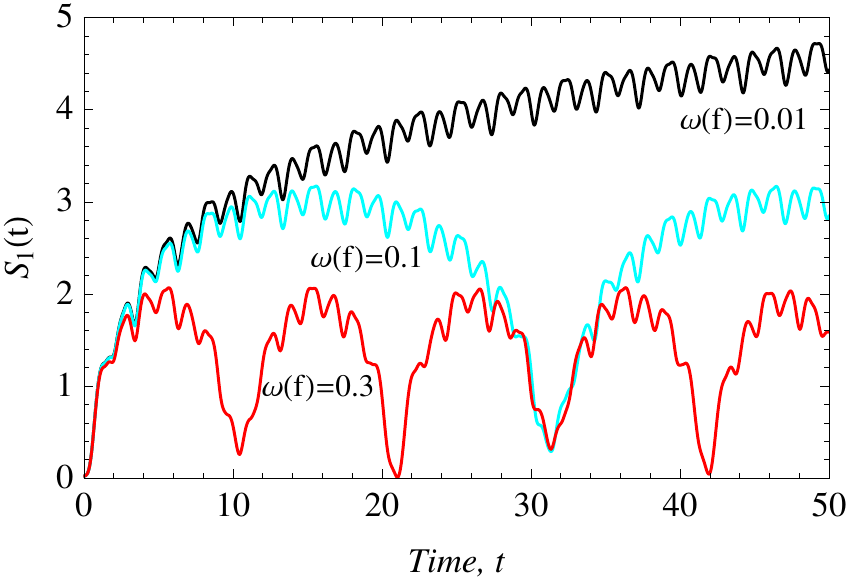}
\caption{Plots of von Neumann entropy for $N=4$. The initial values are $\omega(i)= 3$ and $k(i) =2$. The quenched values are $\omega(f) = 0.3, 0.1, 0.01$ and $k(f) = 2.5$. Three independent time scales contribute to the entanglement dynamics. This plot shows entanglement revival whose time period increases with decreasing $\omega(f)$. Each revival period contains several quasi-revivals on shorter time scales due to the effect of the Ermakov solutions}.
\end{figure}

In order to demonstrate the basic features of the time dependence, we first consider the a chain of $N=4$ oscillators and perform a sudden quench at time $t=0$, when $\omega, k$  change from a constant values $(\omega(i), k(i))$ to  $(\omega(f), k(f))$. The reduced system is defined by tracing out the last two oscillators in the chain. The solutions of the  Ermakov equations with $b_j(t=0) =1$ and $\dot{b}_j(t=0) = 0$ are given by $b_j(t)=\sqrt{n_j\cos(2\sqrt{\lambda_j(f)}t)+m_j}$
where $n_j=\frac{\lambda_j(f)-\lambda_j(i)}{2\lambda_j(f)}$,  $m_j=\frac{\lambda_j(f)+\lambda_j(i)}{2\lambda_j(f)}$ and $\lambda_j(i),$ $\lambda_j(f)$ are the eigenvalues of $K$ before and after the quench. Note that for $N=4$, there are only three distinct eigenvalues of $K$ as 
$\lambda_1=\lambda_3$ with $b_1(t)=b_3(t)$. Thus for $N=4$, there would only be three time scales contributing to the entanglement dynamics.


The results for the von Neumann entropy for $N=4$ under several different quenches are shown in Fig. 2. At large times, the profile of the entanglement dynamics is dominated by the smallest frequency, which being independent of interaction $k$ could be a robust experimental probe for testing the entanglement revival in the harmonic chains. The revival time period increases with decreasing value of the quenched frequency $\omega(f)$. The phenomena of revival has been observed in the entanglement negativity for the non-equilibrium dynamics of harmonic chains \cite{eisler-z}. The important difference in our analysis is the existence of dynamically generated multiple time scales within each revival period, which is a new feature due to the solutions of the Ermakov equations. Such quasi-revivals at shorter time scales encode the effect of the interaction $k$.

\begin{figure}[ht!]
\centering
\includegraphics[width=8.6cm]{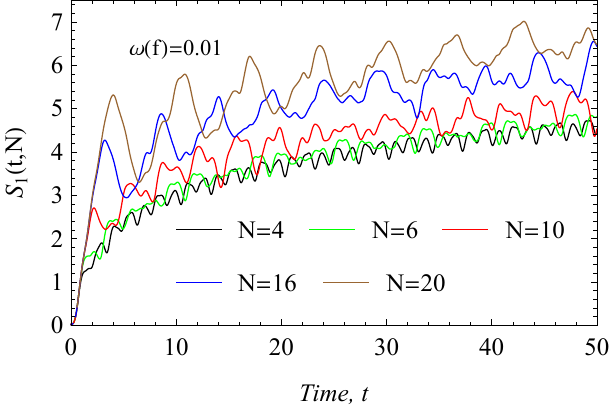}
\label{fig3}
\caption{Plots of von Neumann entropy for $N=4,6,10,16$ and 20. The parameters are $\omega(i)= 3$, $k(i) =2$,  $\omega(f) = 0.01$ and $k(f) = 2.5$. For $N=4,6$ the chains are away from criticality and their entropies are approximately equal, with variations arising from the multiple time scales. For larger $N=10,16,20$, apart from the time dependent factors, even the $N$ dependence makes the plots different, signalling a violation of the area law}.
\end{figure}


 \noindent
  {\it{Scaling of Entropy for large $N$}}- The von Neumann entropy $S_1(t,N)$ is plotted in Fig. 3 as a function of time $t$ for various $N$ with same $\omega(f) = 0.01$. For each value of $N$, the time evolution of $S_1(t,N)$ shows the effect of multiple time scales whose number increases with $N$. In addition, the von Neumann entropy itself increases as a function of $N$. In order to extract the $N$ dependence of the entropy, in Fig. 4 we have plotted the ratio $\frac{S_1(t,N)}{\ln N}$ as a function of time $t$ for several $N$. For $N \geq 10$, the nature of this plot is consistent with the scaling relation 
\begin{equation}
\label{ent-scaling}
S_1 (t,N) = c(t) \ln N + { O}\left (1 \right ), 
\end{equation}
where  $c(t)$ is a time dependent function that encodes the cumulative effect of the dynamically generated multiple time scales.
\begin{figure}[ht!]
\centering
\includegraphics[width=8.6cm]{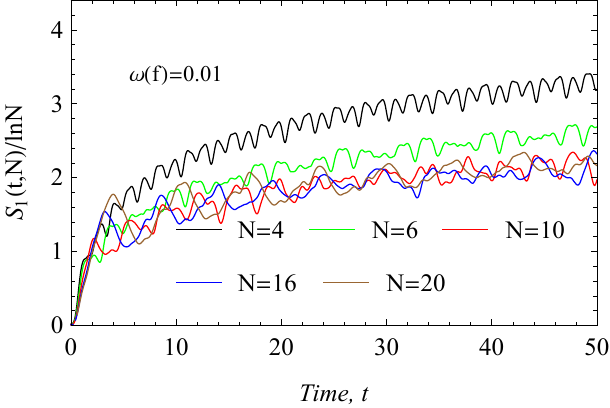}
\label{fig4}
\caption{The entropy curves for $N=10,16$ and 20 almost collapse on each other consistent with Eqn. (\ref{ent-scaling}), the difference being due to the multiple oscillatory time scales. Chains with $N=4$ and 6 are further away from criticality, which is valid only for larger values of $N$}.
\end{figure}
In order to understand the origin of the scaling, note that for $N \rightarrow \infty$ and in the continuum limit, the oscillator chain given by Hamiltonian (\ref{H}) can be described by a single bosonic massive scalar field whose mass is given by $\omega$ \cite{tonni}. Under these conditions, in the limit $\omega \rightarrow 0$, the system tends to criticality leading to a logarithmic scaling of the entropy \cite{wilczek}. In our case, finite $N$ and small but non-zero $\omega(f)=0.01$ makes the system approximately critical. From Fig. 4 we find that the ratio $\frac{S_1(t,N)}{\ln N}$ for $N \geq 10$ essentially collapse on each other, the variations appearing primarily from the time dependent factor $c(t)$, which encodes the effect of the multiple time scales. The function $c(t)$ would in general have contributions from $b_j(t)$ and their time derivatives, all of which are periodic functions. At large time scales, these large number of different periodic functions would tend to produce a smoother time dependence.

From Figs. 3 and 4, we can infer about the validity of area law for the various chains.  For $N\geq 10$, the von Neumann entropy scales as $\ln N$ and does not saturate to any finite fixed value. This indicates a logarithmic violation for these larger chains, which is consistent with their approximate criticality in the post quench regime. The smaller chains are further away from criticality and their entropies essentially coincide upto variations arising from the Ermakov time scales, which is approximately consistent with the area law \cite{eisert}.

\section{Summary and outlook}

We have obtained exact analytical expressions for the time-dependent von Neumann and R\'enyi entropies for a system of $N$ coupled oscillators, following a sudden quench. Our analysis employs the solutions of the TDSE, which are obtained by solving the corresponding nonlinear Ermakov equations. The entanglement dynamics is characterized by a multi-oscillatory behaviour and the number of time scales appearing in the entanglement dynamics increases with $N$. 

The exact analysis and the formulae presented here are valid for any $N$. Although we have used sudden quench to study the entanglement dynamics, the formalism developed here can be easily adapted to investigate the effects of more general time dependence and to study other entanglement measures. 

In the critical limit of our system, we expect the logarithmic scaling of the entropy \cite{wilczek} to be robust against small polynomial perturbations. In addition, the extension of our analysis to finite temperatures using conformal field theory \cite{korepin} would be interesting.

\begin{acknowledgments}
 KSG would like to thank Rodrigo Pereira for discussions and IIP, Natal, RN, Brazil for kind hospitality, where a part of this work was done. The authors thank Dr. S. R. Jain for useful comments. This work is supported by the Department of Atomic Energy (DAE), India.
\end{acknowledgments}

\bibliography{ent1}

\end{document}